\documentclass[final,3p,times,twocolumn]{elsarticle}

\usepackage{amssymb}
\usepackage{amsthm}
\usepackage{amsmath}
\usepackage{pifont}
\usepackage{hyperref}

\usepackage{lineno}

\usepackage{multirow}
\usepackage{siunitx}

\journal{NIM A}

\begin{document}

\begin{frontmatter}



\title{Performance of CVD diamond detectors for single ion beam-tagging applications in hadrontherapy monitoring}


\author[LPSC]{S. Curtoni\corref{SC}}
\author[LPSC]{M.-L. Gallin-Martel}
\author[LPSC]{S. Marcatili}
\author[NEEL]{L. Abbassi}
\author[LPSC]{A. Bes}
\author[LPSC]{G. Bosson}
\author[LPSC]{J. Collot}
\author[NEEL]{T. Crozes}
\author[LPSC]{D. Dauvergne}
\author[ESRF]{W. De Nolf}
\author[LPSC]{P. Everaere}
\author[LPSC]{L. Gallin-Martel}
\author[LPSC]{A. Ghimouz}
\author[ARRONAX,SUBATECH]{F. Haddad}
\author[LPSC]{C. Hoarau}
\author[LPSC]{J.-Y. Hostachy}
\author[ARRONAX,SUBATECH]{C. Koumeir}
\author[LPSC]{A. Lacoste}
\author[SUBATECH]{V. M\'etivier}
\author[ESRF]{J. Morse}
\author[NEEL]{J.-F. Motte}
\author[LPSC]{J.-F. Muraz}
\author[ARRONAX,SUBATECH]{F. Poirier}
\author[LPSC]{F. E. Rarbi}
\author[LPSC]{O. Rossetto}
\author[ESRF]{M. Salom\'e}
\author[SUBATECH]{N. Servagent}
\author[IP2I]{E. Testa}
\author[LPSC]{M. Yamouni}

\address[LPSC]{Universit\'e Grenoble-Alpes, CNRS, Grenoble INP, LPSC-IN2P3 UMR 5821, 38000 Grenoble, France}
\address[NEEL]{Universit\'e Grenoble-Alpes, CNRS, Institut N\'eel, NANOFAB UPR2940, 38000 Grenoble, France}
\address[ESRF]{European Synchrotron Radiation Facility, 38000 Grenoble, France}
\address[ARRONAX]{GIP ARRONAX, 44800 Saint Herblain, France}
\address[SUBATECH]{Universit\'e de Nantes, CNRS, IMT Atlantique, SUBATECH-IN2P3 UMR 6457, 44000 Nantes, France}
\address[IP2I]{Universit\'e de Lyon, CNRS, IP2I-IN2P3 UMR 5822, 69000 Lyon, France}

\cortext[SC]{Corresponding author: curtoni@cppm.in2p3.fr}

\begin{abstract}

In the context of online ion range verification in particle therapy, the CLaRyS collaboration is developing Prompt-Gamma (PG) detection systems. The originality in the CLaRyS approach is to use a beam-tagging hodoscope in coincidence with the gamma detectors to provide both temporal and spatial information of the incoming ions. The ion range sensitivity of such PG detection systems could be improved by detecting single ions with a \SI{100}{\pico\second} ($\sigma$) time resolution, through a quality assurance procedure at low beam intensity at the beginning of the treatment session. This work presents the investigations that led to assessment of the Chemical Vapor Deposition (CVD) diamond detectors performance to fulfil these requirements. A $^{90}$Sr beta source, \SI{68}{\mega\electronvolt} protons, 95 MeV/u carbon ions and a synchrotron X-ray pulsed beam were used to measure the time resolution, single ion detection efficiency and proton counting capability of various CVD diamond samples. An offline technique,  based on double-sided readout with fast current preamplifiers used to improve the signal-to-noise ratio, is also presented. The different tests highlighted Time-Of-Flight resolutions ranging from \SI{13}{\pico\second} ($\sigma$) to \SI{250}{\pico\second} ($\sigma$), depending on the diamond crystal quality and the particle type and energy. The single \SI{68}{\mega\electronvolt} proton detection efficiency of various large area polycrystalline (pCVD) samples was measured to be $>$96\% using coincidence measurements with a single-crystal reference detector. Single-crystal CVD (sCVD) diamond proved to be able to count a discrete number of simultaneous protons while it was not achievable with a polycrystalline sample. Considering the results of the present study, two diamond hodoscope demonstrators are under development: one based on sCVD, and one of larger size based on pCVD. They will be used for the purpose of single ion as well as ion bunches detection, either at reduced or clinical beam intensities.

\end{abstract}



\begin{keyword}
CVD diamond \sep hadrontherapy \sep ion range verification \sep time resolution \sep detection efficiency \sep particle counting \sep beam monitoring



\end{keyword}

\end{frontmatter}



\section{Introduction}
\label{sect:Intro}
Hadrontherapy is an external radiotherapy modality based on light ion beams \citep{Newhauser2015, Schardt2010}. Even though the ballistic properties of ions and the enhanced relative biological effectiveness represent essential advantages of particle therapy compared to conventional X-ray radiotherapy, it is still facing limitations due to ion range uncertainties arising at every stage of the treatment procedure \citep{Paganetti2012}. They currently lead physicians to set ion range specific safety margins that limit the dose conformation and prevent them to plan irradiation fields where organs at risk are located close beyond the targeted volume.

In this context, several experimental approaches have been developed to build an online ion range verification system \citep{Knopf2013, Kraan2015}. Among them, prompt-gamma-based verification techniques \citep{Krimmer2018} propose to retrieve the actual ion range from the emission profile of prompt-gamma photons (PG) that are emitted along the ion path by excited target nuclei or ion fragments right after inelastic collisions between incoming ions and target nuclei. To get rid of the inherent and substantial neutron-induced background also produced during these nuclear interactions, PG detection systems use a Time-Of-Flight (TOF) based gamma-neutron discrimination. It is generally carried out by coincidence measurements between the gamma camera trigger and the ion bunch time of arrival given by the accelerator radio-frequency signal (RF). Provided the body-camera distance is set to a few tens of centimeters, an overall TOF resolution of \SI{1}{\nano\second} ($\sigma$) is sufficient to achieve this purpose.

Instead of using the accelerator RF as a START signal, the CLaRyS collaboration proposes to set up a beam-tagging hodoscope upstream from the patient at reduced intensity ($\sim$ 1 ion/bunch). It will also provide an ion transverse position that is useful for PG vertices reconstruction with PG imaging systems (PGI). The direct detection of incoming ions also makes the TOF measurement independent of the beam time structure and/or any potential RF phase shift as has been observed \citep{Werner2019}. According to this idea, the collaboration has developed a $12.8 \times \SI{12.8}{\square\centi\metre}$ scintillating-fiber hodoscope. It has been tested and characterized on proton and carbon ion beams \citep{Allegrini2020} and the results highlighted a \SI{0.7}{\nano\second} ($\sigma$) time resolution and a detection efficiency up to 98\%.

Considerable improvements can be achieved in the sensitivity of potential ion range shift determination by improving the TOF resolution down to a few hundred picoseconds. This holds for PGI and prompt gamma timing (PGT) \citep{Golnik2014, Hueso-Gonzalez2015, Marcatili2020} and is thoroughly discussed in \citep{Dauvergne2020}. Different detector technologies could enable the development of a beam monitor with a 100 ps ($\sigma$) time resolution for single ion detection \citep{Vignati2017, Federici2020, BossiniMinafra2020}. The collaboration has chosen to focus on Chemical Vapor Desposition (CVD) diamond technology in order to develop a beam hodoscope upgrade combining an excellent time resolution \citep{Pomorski2006},\citep{BossiniMinafra2020} (and references therein) and high radiation hardness guaranteeing long-term stability in clinical conditions.

The current work presents investigations led on diamond detectors, at first, to evaluate polycrystalline (pCVD) single proton detection efficiency. Then, experiments were carried out to assess single crystal (sCVD), pCVD and Diamond On Iridium (DOI) detector ability to perform TOF measurements with a 100 ps resolution, using 68 MeV single protons in ARRONAX (Saint-Herblain, France), 95 MeV/u carbon ions in GANIL (Caen, France), short pulses of 8.53 keV X-rays at ESRF (Grenoble, France) and minimum ionizing particle (MIP) with a $^{90}$Sr laboratory source. Finally, the sCVD and pCVD diamond detectors single particle counting capabilities have been evaluated with the 68 MeV proton beam delivered by the ARRONAX cyclotron at low intensity (6 pA $\sim$ 1 proton/bunch).


\section{Material and methods}
\subsection{Detectors assembly and generic experimental set-up}

The detector-grade diamond samples used in the present work are commercially available and produced by Chemical Vapor Deposition (CVD). The sCVD diamonds were purchased from Element6 \citep{Element6}, pCVD diamonds from Element6, II-VI \citep{II-VI} and Diamond Delaware Knives (DDK) \citep{DDK}, and DOI diamonds from Audiatec \citep{Audiatec} and Augsburg University. The tested samples ranged from 300 \si{\micro\metre} to 500 \si{\micro\metre} in thickness, and from 4.5 $\times$ 4.5 \si{\square\milli\metre} to 20 $\times$ 20 \si{\square\milli\metre} in area. In particular, large pCVD are foreseen for the assembly of a large size hodoscope. Diamond samples were assembled as pad detectors as described in \citep{Gallin-Martel2016, Gallin-Martel2018}. A thin aluminum disk-shaped metallization was performed either by physical evaporation \citep{Lacoste2002} (50 nm) or by sputtering (100 nm). The diamonds were sandwiched between two 50\si{\ohm}-adapted printed circuit boards (PCB), allowing bias of both polarities and signal readout connections on both sides.

\begin{table*}[t!]
\centering
\caption{Summary of the various diamond samples tested within this work. DE = Single proton detection efficiency, TOF = Time-Of-Flight resolution, C = counting, $\sigma_t$ = intrinsic time resolution.}
\label{tab:all_diamonds}
\resizebox{\textwidth}{!}{%
\begin{tabular}{|c|c|c|c|c|c|c|c|}
\hline
\multirow{2}{*}{Diamond} &
  \multirow{2}{*}{Provider} &
  \multirow{2}{*}{\begin{tabular}[c]{@{}c@{}}Size\\ ($\si{\cubic\milli\metre}$)\end{tabular}} &
  \multicolumn{2}{c|}{Metallization} &
  \multirow{2}{*}{\begin{tabular}[c]{@{}c@{}}Computed\\ capacitance (pF)\end{tabular}} &
  \multirow{2}{*}{Tested with} &
  \multirow{2}{*}{\begin{tabular}[c]{@{}c@{}}Involved in\\ measurements:\end{tabular}} \\ \cline{4-5}
 &
   &
   &
  Diam. (mm) &
  Thick. (nm) &
   &
   &
   \\ \hline
sCVD &
  E6 &
  4.5 $\times$ 4.5 $\times$ 0.517 &
  3 &
  50 &
  0.7 &
  X, $\beta$, p, $^{12}$C &
  DE, TOF, C \\ \hline
sCVD &
  E6 &
  4.5 $\times$ 4.5 $\times$ 0.517 &
  3 &
  50 &
  0.7 &
  $\beta$ &
  TOF \\ \hline
pCVD &
  E6 &
  10 $\times$ 10 $\times$ 0.3 &
  7 &
  50 - 100 &
  6.5 &
  X, p, $^{12}$C &
  DE, TOF \\ \hline
pCVD &
  E6 &
  20 $\times$ 20 $\times$ 0.5 &
  16 &
  50 &
  20 &
  $^{12}$C &
  TOF \\ \hline
pCVD &
  II-VI &
  10 $\times$ 10 $\times$ 0.5 &
  7 &
  100 &
  3.9 &
  p &
  DE, TOF \\ \hline
pCVD &
  DDK &
  5 $\times$ 5 $\times$ 0.3 &
  3 &
  100 &
  1.2 &
  p &
  DE, TOF, C \\ \hline
DOI &
  \begin{tabular}[c]{@{}c@{}}Augsburg\\ Univ.\end{tabular} &
  5 $\times$ 5 $\times$ 0.3 &
  3 &
  50 - 100 &
  1.2 &
  X, p, $^{12}$C &
  TOF, $\sigma_t$ \\ \hline
DOI &
  Audiatec &
  5 $\times$ 5 $\times$ 0.3 &
  3 &
  50 &
  1.2 &
  X &
 $\sigma_t$ \\ \hline
\end{tabular}%
}
\end{table*}

\begin{figure}[h!]
    \begin{center}
    \includegraphics[width=1.0\columnwidth]{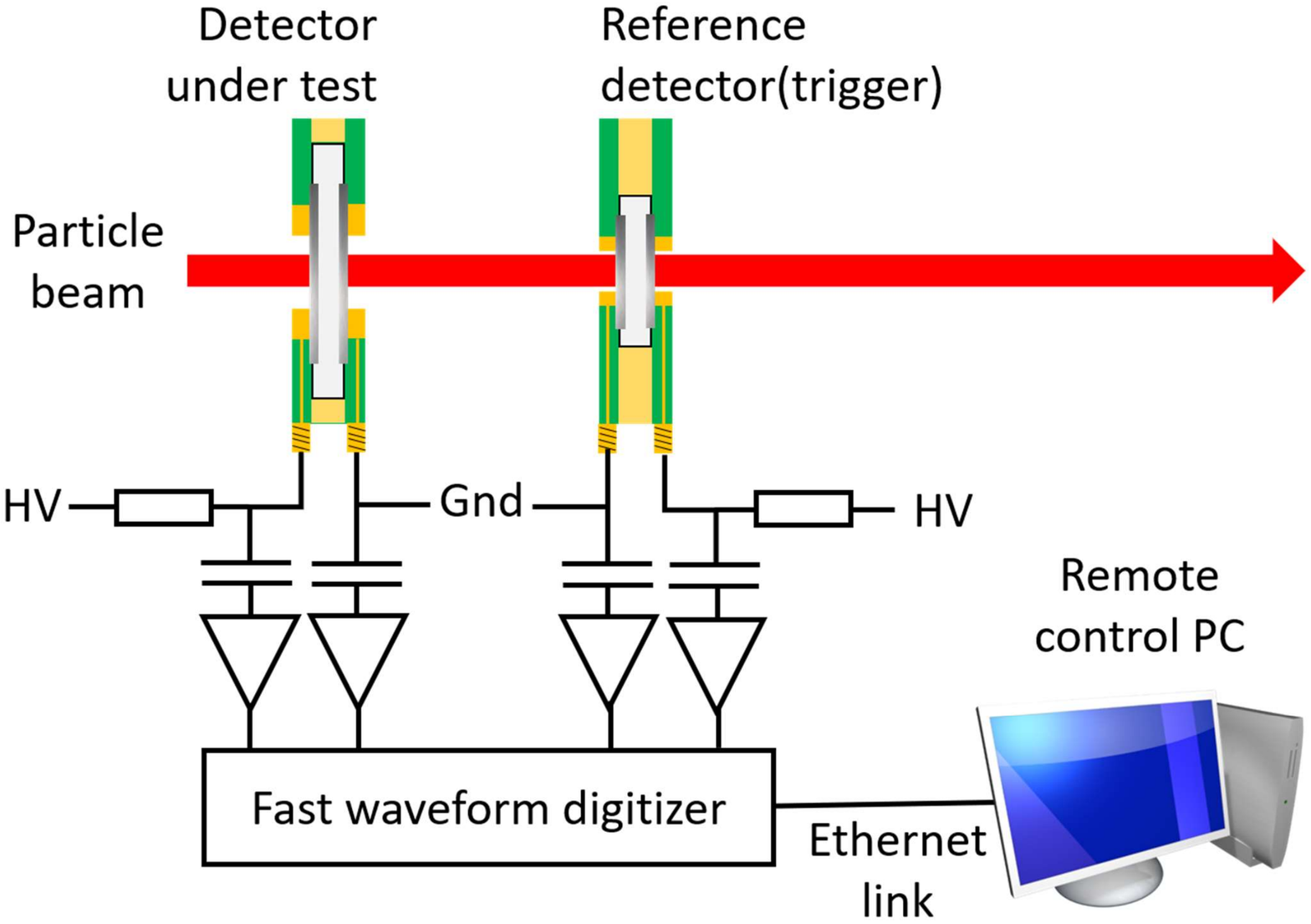}
    \caption{Generic experimental set-up used for detection efficiency, timing and counting measurements presented in this work. Specific dedicated additions in the set-up are presented in the corresponding subsections.}
    \label{fig:schema_timing_setup}
    \end{center}
\end{figure}

For the different tests presented in this work, diamond detectors were systematically tested by pair. Figure \ref{fig:schema_timing_setup} illustrates the general configuration used during the tests. Two diamond detectors are exposed to a particle beam. The detector under test is positioned upstream from a smaller size reference detector (a sCVD sample, unless stated otherwise). For each beam test, both detectors were enclosed together in an aluminum shielding box with front and rear apertures covered with \SI{12}{\micro\metre}-thick aluminized Mylar films \citep{Gallin-Martel2016, Gallin-Martel2018}. The output channels of the detectors were coupled to broadband amplifiers (CIVIDEC C2-HV \citep{Cividec} or Greenstream DBA III and IV-R) and analog signals were digitized using fast digitizers (a 500 MHz, 3.2 GS/s WaveCatcher digitizer \citep{Breton2015} or a 2 GHz, 20 GS/s LeCroy Digital Storage Oscilloscope (DSO)).

\subsection{Side-to-side signals summation}
\label{subsect:matmeth_summing}
During the interaction of a ionizing particle in a diamond detector, the same signal $S$ (absolute value) is induced on both electrodes by the electron/hole pairs drift. In practice, each side is read by a single preamplifier generating an output signal of amplitude $S^{side_i}$ (with i=1,2) with a corresponding noise level $\sigma_{n}^{side_i}$ resulting in a signal-to-noise ratio $S^{side_i}/\sigma_{n}^{side_i}$. First, we assume that the intrinsic noise generated by the diamond itself is negligible at \SI{300}{\kelvin} compared to that induced by the wide-band preamplifier. Then, the noise of each preamplifier is assumed to be an independent Gaussian white noise. Therefore, the resulting noise on the sum-signal $\sigma^{sum}_n$ can be expressed as follows: 

\begin{equation}
\label{eq:N_somme_general}
\sigma_{n}^{sum} = \sqrt{\left(\sigma_{n}^{side_{1}}\right)^2 + \left(\sigma_{n}^{side_{2}}\right)^2} = \sigma_{n}^{side_{1}} \oplus \sigma_{n}^{side_{2}}\,.
\end{equation}

\noindent
A sum signal $S^{sum} = S^{side_1}-S^{side_2}$ (the two side signals are of opposite polarity) can be derived as well as a sum-signal-to-noise ratio:

\begin{equation}
\label{eq:SNR_somme_general}
S/N_{sum} = \frac{S^{side_{1}} - S^{side_{2}}}{\sigma_{n}^{side_{1}} \oplus \sigma_{n}^{side_{2}}}\,.
\end{equation}

\noindent
If one supposes now that the two preamplifiers are strictly identical, Equation \ref{eq:SNR_somme_general} becomes:

\begin{equation}
\label{eq:SNR_somme}
S/N_{sum} = \frac{2S^{side}}{\sqrt{2\sigma_{n}^2}} = \frac{2S^{side}}{\sigma_{n}\sqrt{2}} = \sqrt{2} \cdot S/N_{side}\,.
\end{equation}

\noindent
Using the sum signal, the signal-to-noise ratio (SNR) can be increased by a factor $\sqrt{2}$. By doubling the amplitude of the signal, the slope in the rising edge of the sum-signal can become up to twice as much as the one measured on each electrode signal improving the time resolution of the detector consequently. This technique was used for the detection efficiency and timing measurements presented in Sections \ref{subsubsec:matmeth_deteff} and \ref{subsubsec:matmeth_timeres}. Note that this technique requires identical preamplifiers (same pulse shape), a strict adjustment of the pulses risetime and a null delay between both side signals.

\subsection{Experimental tests and data analysis procedures}

\subsubsection{Single proton detection efficiency of pCVD detectors}
\label{subsubsec:matmeth_deteff}

The single proton detection efficiency has been evaluated with 68 MeV protons during a dedicated experiment at ARRONAX IBA C70 isochronous cyclotron, Nantes (with a fixed Radio-Frequency of 30.45 MHz) \citep{Poirier2011, Poirier2016}. In order to restrict the incoming beam to bunches containing at most one single proton, the beam intensity was lowered down to 50 fA. Three different pCVD detectors presented in Table \ref{tab:all_diamonds} and based on samples coming from different providers were tested one-by-one during the experiment in reproducible conditions. Each pCVD sample was tested in coincidence with the same sCVD reference detector. The detectors box was set up and aligned between two 2.5 cm-thick aluminum collimators with \SI{5}{\milli\metre} gaps. The upstream one was used to constrain the proton incidence to the sensitive surface of the smallest detector. The downstream one reduced the beam halo caused by the scattering of protons in the PCBs. Behind the second collimator, a PTW T34058 gas ionization chamber (IC) and a \SI{5}{\milli\metre}-thick plastic scintillator coupled to a photomultiplier tube (PMT) were aligned with the beam.  The IC was coupled to a PTW Unidos electrometer to measure the beam current while the scintillator was used to get a redundant spectroscopic information of incoming ions that was used for the efficiency measurement. The applied bias voltage was +300 V for the Element6 and DDK pCVD detectors, +500V for the II-VI pCVD detector and +300 V for the sCVD detector, according to the scheme presented in Figure \ref{fig:schema_timing_setup}. The applied biasing was +400 V for the IC and -800 V for the PMT. The two pCVD output channels were coupled to Greenstream DBA IV-R preamplifiers \citep{Moritz2001} while CIVIDEC C2-HV preamplifiers were used for the sCVD sample. Analog signals from the diamond detectors and the scintillator were sampled using the WaveCatcher digitizer. 

To assess the single proton detection efficiency of pCVD samples, measurements in coincidence with the two reference detectors (the sCVD and the scintillator) were used. First, recorded events that corresponded to a double coincidence between the two reference detectors were identified using a coincidence window of duration $\delta t = \SI{1.25}{\nano\second}$, as well as low and high voltage thresholds selecting single proton events. Among the $N_{double}(\delta t)$ events that corresponded to these criteria, triple and random coincidences were tested event-by-event on the pCVD samples using two coincidence windows and a voltage threshold scanning. The triple coincidence window was the same as the one applied on the reference detectors. The random coincidence window was delayed by \SI{15}{\nano\second}, between two consecutive bunches (\SI{32.84}{\nano\second}). Using the voltage threshold sweep, a voltage comparison is performed on the pCVD signal between the threshold level $V_{th}$ and the waveform segments contained within the coincidence windows. For each $V_{th}$ value, we counted $N_{triple}\left(V_{th}; \delta t\right)$ triple coincidences between the pCVD and the two reference detectors and $N_{random}\left(V_{th}; \delta t\right)$ random coincidences triggered by noise fluctuations in the pCVD signals. Thus, we can define a true coincidence detection efficiency $\epsilon\left(V_{th}; \delta t\right)$ at a given threshold value $V_{th}$ ($\delta t$ is a fixed parameter) as follows:

\begin{equation}
\label{eq:det_eff}
\epsilon\left(V_{th}; \delta t\right) = \frac{N_{triple}\left(V_{th}; \delta t\right)}{N_{double}\left(\delta t\right)} \times \left(1 - \frac{N_{random}\left(V_{th}; \delta t\right)}{N_{double}\left(\delta t\right)}\right) \,.
\end{equation}

Equation \ref{eq:det_eff} can be understood as the product of the probability to detect a true triple coincidence and the probability not to detect a random coincidence. As $\delta t$ is fixed, the single proton detection efficiency $\epsilon_{det}$ was here defined as the maximum of the obtained $\epsilon(V_{th})$ function.

\subsubsection{Timing resolution with single ions}
\label{subsubsec:matmeth_timeres}

The experimental set-up presented in Figure \ref{fig:schema_timing_setup} was used in various beam test configurations as listed in Section \ref{sect:Intro} in order to evaluate the TOF resolution achievable between two diamond detectors of various crystalline qualities. Within the scope of this article, the TOF resolution $\sigma_{TOF}$ of a pair of independent detectors with respective time resolution $\sigma_{t_1}$ and $\sigma_{t_2}$ is defined as:

\begin{equation}
    \sigma_{TOF} = \sqrt{\sigma_{t_1}^2 + \sigma_{t_2}^2} \,.
\end{equation}

Timing measurements were carried out on the digitized waveforms using a normalised threshold algorithm, as defined in \citep{Berretti2015}. Once the amplitude of a pulse is detected, a constant fraction of this value (between 20\% and 50\%) is computed. The pulse time stamp is finally obtained by means of a linear interpolation between waveform samples, thus emulating an analog Constant Fraction Discrimination (CFD). Unless stated otherwise, the timing resolution is derived from the statistical dispersion measured on the timing difference between the sum signals of the two detectors involved. Furthermore, a pair of diamond detectors composed of a $5.0 \times 5.0 \times \SI{0.3}{\cubic\milli\metre}$ DOI detector produced at Augsburg University and a $4.5 \times 4.5 \times \SI{0.517}{\cubic\milli\metre}$ sCVD detector produced by Element6 were tested together in the different beam tests presented in this work. They were systematically tested to provide a common reference for comparison purposes and are referred as the sCVD-DOI reference pair, later on in this article.

At ARRONAX, the timing measurements were carried out on the waveform datasets we acquired for the single proton detection efficiency assessment. We could therefore measure the TOF resolution for the three pCVD-sCVD couples, as they are presented in Table \ref{tab:all_diamonds} and Section \ref{subsubsec:matmeth_deteff}. For each $V_{th}$ value, the events subset which fulfilled the triple coincidence criterion and did not trigger random coincidences was selected. On this subset, the pulse discrimination was performed using the normalised threshold algorithm on the pCVD and sCVD sum signals. The distribution of the timing difference between the discriminated pCVD and sCVD signals was then stored in a histogram. Since some distributions demonstrated non-gaussian tails on each side, the root mean square (RMS) value of the histogram was chosen as an estimator of the TOF resolution for all histograms (\textit{i.e.} for each $V_{th}$ value).

The timing measurements at GANIL were carried out with single 95 MeV/u carbon ions and the standard bench presented in Figure \ref{fig:schema_timing_setup}. Yet, one noteworthy difference is that only one CIVIDEC C2-HV preamplifier could be used for each diamond detector during this test, preventing us from using the signal summation technique introduced in Section \ref{subsect:matmeth_summing}. The waveforms were digitized with the 3.2 GS/s WaveCatcher system. Two pairs of detectors were tested. The first one is the sCVD-DOI reference pair and the second one is composed of two Element6 pCVD detectors, $20 \times 20 \times \SI{0.5}{\cubic\milli\metre}$ and $10 \times 10 \times \SI{0.3}{\cubic\milli\metre}$ respectively. They were metallized as pad detectors, with disk-shaped 50nm-thick Al electrodes and respective diameter of \SI{16}{\milli\metre} and \SI{7}{\milli\metre}.

\subsubsection{Timing resolution with a pulsed X-ray beam}

At ESRF, the combined use of a X-ray micro-beam and various attenuators set up upstream from the detectors enabled us to study the time response of a pair of diamond detectors as a function of the energy deposition. The test beam took place in ID21 beamline \citep{Cotte2017} that delivered a \SI{8.53}{\kilo\electronvolt} X-ray micro-beam while the ESRF synchrotron was running in 4-bunch mode. In this configuration, the pulsed beam RF was $f_{RF} = \SI{1.42}{\mega\hertz}$ ($T_{RF} = \SI{704}{\nano\second}$) and the bunch duration was \SI{100}{\pico\second}. With the maximum electron beam current (\SI{32}{\milli\ampere}) circulating in the synchrotron, the primary X-ray flux was $\phi_{32mA} = 1.79 \cdot 10^9$ photons/s \citep{GallinMartel2020}, which corresponds to $1.26 \cdot 10^3$ photons/bunch. The absorption length of X-rays with an energy $E_X = \SI{8.53}{\kilo\electronvolt}$ in diamond is $1/\mu_{diam} \sim$ 790 \si{\micro\metre} \citep{XCOM}. As a result, the energy deposition is almost uniformly distributed over the thickness of the tested samples (300 - 500 \si{\micro\metre}) thus mimicking the passage of single charged particles.

The two detectors used here were a $4.5 \times 4.5 \times \SI{0.517}{\cubic\milli\metre}$  Element6 sCVD detector and a $5.0 \times 5.0 \times \SI{0.3}{\cubic\milli\metre}$ Audiatec DOI detector. Both were metallized with aluminium disk electrodes of 3 mm diameter. For each attenuator used (Al and Ti foils with various thicknesses), an acquisition of the signals coming from the two detectors as well as of the RF signal was performed. Each side electrode of the Audiatec sensor was coupled to a CIVIDEC C2-HV preamplifier while only one was used on the sCVD detector. For a given attenuator type and thickness, the energy deposits of a X-ray bunch in the DOI and sCVD detectors (thereafter noted $\Delta E_{DOI}$ and $\Delta E_{sCVD}$) are computed using Beer-Lambert law as follows :

\begin{equation}
\begin{split}
    \Delta E_{DOI} &= E_{X} \cdot \frac{\phi_{32mA}}{f_{RF}} \cdot \exp{(-\mu_{att}x_{att}-\mu_{PET}x_{PET})}\\
    &\cdot \left[1-\exp{(-\mu_{diam}d_{DOI})}\right] \,,
    \label{eq:deltaE_DOI}
\end{split}
\end{equation}

\begin{equation}
\begin{split}
        \Delta E_{sCVD} &= E_{X} \cdot \frac{\phi_{32mA}}{f_{RF}} \\
        &\cdot \exp{(-\mu_{att}x_{att}-\mu_{PET}x_{PET}-\mu_{diam}d_{DOI})} \\
        &\cdot \left[1-\exp{(-\mu_{diam}d_{sCVD})}\right] \,,
        \label{eq:deltaE_mono}
\end{split}
\end{equation}

\noindent
where $\mu$ and $x$ are respectively the attenuation coefficient at 8.53 keV and the thickness of the considered material (att = attenuator, PET = Mylar, diam = diamond) while $d$ is the thickness of the detector. In this set-up, the attenuation ($< 0.7\%$) of the beam in the air path between the detectors was neglected. For each acquisition, the signals have been processed using the normalised threshold algorithm at 50\%, as defined in Section \ref{subsubsec:matmeth_timeres}. 

A previous experiment with the DOI detector from Augsburg University had been carried out with only a few number of attenuators. During this test, the Augsburg DOI sample was coupled to two preamplifiers while the sCVD detector was equipped to only one preamplifier. In this case, we had measured the side-to-side time difference between pulses generated on the two electrodes of the DOI detector. We performed the same measurement with the Audiatec sample and compared the timing performance obtained in both cases.

\subsubsection{Timing resolution with MIP electrons}

Using MIP-like electrons allowed us to determine a lower bound of the timing resolution that could be obtained for the detection of single particles. In this case, two Element6 sCVD diamond detectors (4.5 $\times$ 4.5 $\times$ 0.51 \si{\cubic\milli\metre} each) were used. They were exposed to a collimated beam of electrons from a $^{90}$Sr source, with an energy up to \SI{2.28}{\mega\electronvolt}. The source, the collimators and the detectors were all enclosed in a U-shaped rail ensuring the mechanical alignment of the set-up. An assembly of four scintillating fibres coupled to a common PMT was added downstream from the diamond detectors. It was used as an external trigger to detect electrons in the higher energy part of the $\beta$ spectrum. Both electrodes of the two diamond detectors were coupled to CIVIDEC C2-HV preamplifiers via 10 cm coaxial cables. The signals produced by the four preamplifiers were digitized using a LeCroy HDO9404 DSO (4 GHz, 20 GS/s, 10 bits). The applied bias voltage was -500 V on the two diamond detectors.

\subsubsection{Proton counting}
\label{subsubsec:matmeth_counting}

\begin{figure}[h!]
    \begin{center}
    \includegraphics[width=1.0\columnwidth]{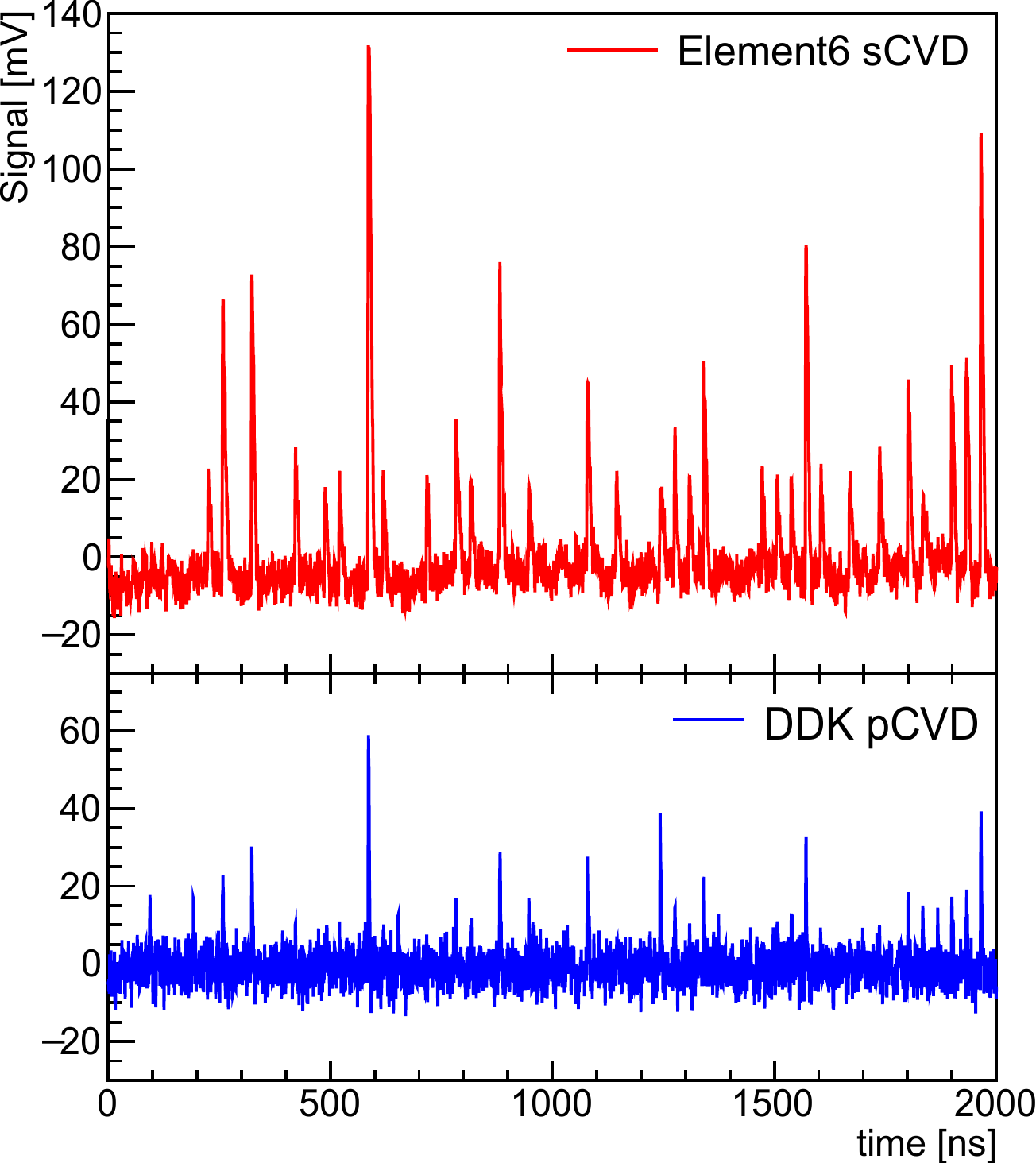}
    \caption{Compared waveforms acquired simultaneously on the Element6 sCVD detector (top, red) and the DDK pCVD detector (bottom, blue) using a 2.5 GS/s sampling rate.}
    \label{fig:WF_publi_final}
    \end{center}
\end{figure}

The counting and monitoring capabilities of the diamond samples were also tested at ARRONAX. The DDK pCVD detector and the Element6 sCVD detector presented in Table \ref{tab:all_diamonds} were selected for this test. Only one output channel per detector was used here and the biased electrodes of the pCVD and the sCVD detectors were coupled to one preamplifier. In order to acquire \SI{2}{\micro\second}-long waveforms (corresponding to 60 RF periods at 30.45 MHz), the sampling rate was lowered down to 2.5 GS/s. A 2.5 cm-thick aluminum collimator with a \SI{1}{\milli\metre} gap was set up in front of the detectors to constrain the beam to a section smaller than the sensitive diameter of the detectors. Typical signal waveforms acquired simultaneously on the two detectors are shown in Figure \ref{fig:WF_publi_final}. 

In order to count the number of protons contained in the bunches, charge measurement was performed by numerical integration of the waveforms on both detectors. First, a baseline correction was achieved by projecting all the waveform samples voltage values in an histogram. Considering that some RF periods do not contain protons at this beam current level ($\sim$\SI{5}{\pico\ampere}) and that the signal duration is short compared to the RF period, the histogram exhibits a dominant Gaussian noise peak that can be fitted to derive its mean and standard deviation parameters. They are then defined as estimators of the baseline offset value and the noise level $\sigma$ of the considered waveform, respectively. After subtraction of the obtained offset, each waveform is subdivided in 60 33ns-long segments (corresponding to the 60 RF periods). For each RF period, the numerical integration is done by summing up the samples contained in the corresponding segment. The charge response of both detectors can thus be compared on a bunch-by-bunch basis, as presented in Section \ref{subsec:results_counting}. 

From the counting statistics, it is possible to derive a mean beam current value. Later on in this paper, we will consider that at a given beam current $I_{beam}$, the number of protons contained in a bunch is a discrete random variable $X$ according to Poisson law, with a $\lambda$ parameter such as $\lambda \propto I_{beam}$. The probability $P(X = k)$ of having $k$ protons in a bunch is therefore:

\begin{equation}
     P(X=k) = \frac{\lambda^{k}}{k!}e^{-\lambda} \,.
     \label{eq:Poisson}
\end{equation}

\noindent
In the case of an ideal beam delivering exactly one proton per bunch ($Q_{bunch} = e$) with a period $T_{beam} = \SI{32.84}{\nano\second}$ (corresponding to the period of the ARRONAX cyclotron RF signal), the average beam current $I_{ref}$ is given by:

\begin{equation}
    I_{ref} = \frac{Q_{bunch}}{T_{beam}} = \frac{1.602 \cdot 10^{-19}}{32.84 \cdot 10^{-9}} = 4.872 \cdot 10^{-12}\si{\ampere\second}/\si{\second} \,.
    \label{eq:current_1p_per_bunch}
\end{equation}

\noindent
Then the average beam current $I_{beam}$ can be derived as follows:

\begin{equation}
    I_{beam} = \lambda I_{ref} \,.
    \label{eq:ibeam}
\end{equation}

\noindent
This expression will be used later on in this work to estimate the average beam current during the counting experiment. Since $\lambda$ is the parameter of the Poisson law describing a bunch's proton multiplicity, this analysis carried out on time windows corresponding to 60 consecutive bunches results in a standard deviation $\sigma_{\lambda} = \sqrt{\lambda/60}$.


\section{Results}
\subsection{Single proton detection efficiency}

\begin{figure}[h!]
    \begin{center}
    \includegraphics[width=1.0\columnwidth]{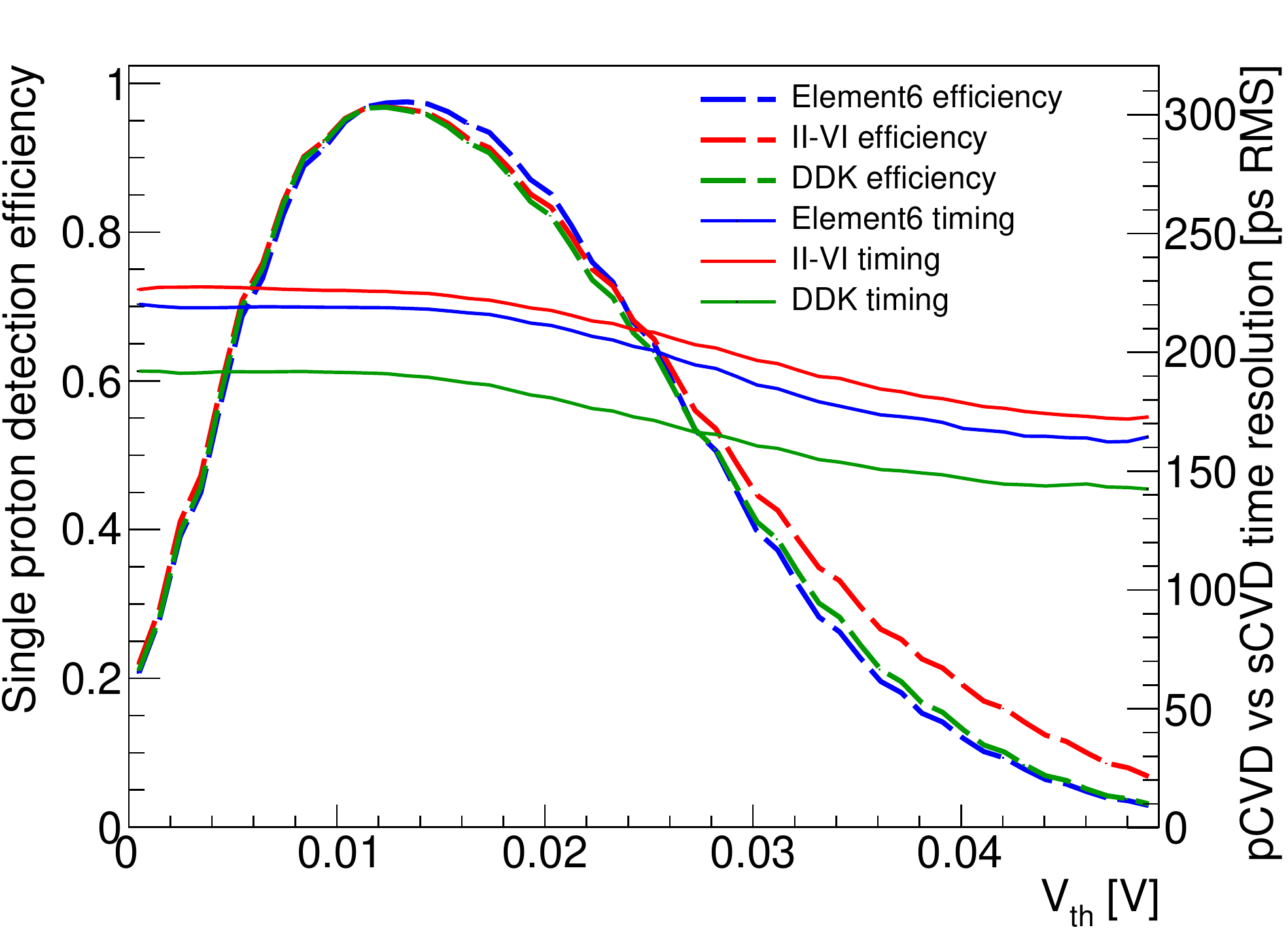}
    \caption{Single 68 MeV proton detection efficiency (dashed lines) and TOF resolution (solid lines) of three pCVD detectors as a function of the threshold value $V_{th}$ used for the pCVD sum signal discrimination. The coincidence window duration is $\delta t = \SI{1.25}{\nano\second}$.}
    \label{fig:ARRONAX2019_eff_timeRes_somme}
    \end{center}
\end{figure}

The results of the analysis presented in Section \ref{subsubsec:matmeth_deteff} and carried out on the sum signals of the three pCVD samples presented on Table \ref{tab:all_diamonds} are combined in Figure \ref{fig:ARRONAX2019_eff_timeRes_somme} (dashed lines). The three pCVD detectors highlight the same behaviour according to the $V_{th}$ value. If $V_{th}$ is close to zero, the probability of a random coincidence triggered by noise fluctuations is comparable to the probability to trigger on the true event pulse. As a consequence, $\epsilon$ remains low. If $V_{th}$ increases, the noise-triggered random coincidence probability decreases and $\epsilon$ increases. Beyond an optimal $V_{th}$ value for which $\epsilon$ is maximized, the threshold starts rejecting true events resulting in the degradation of the detection efficiency.

\noindent
Following Section \ref{subsubsec:matmeth_deteff}, the single \SI{68}{\mega\electronvolt} proton detection efficiency $\epsilon_{det}$ is here defined as $\epsilon_{det} = \max{(\epsilon(V_{th}))}$. For the three pCVD detectors, $\epsilon_{det}$ is obtained at $V_{th} \sim \SI{13}{\milli\volt}$ and reaches 98\% for the Element6 sample, while 97\% is obtained in the case of the II-VI and DDK samples. These results are in good agreement with measurements carried out in similar conditions in a previous study \citep{Frais-Kolbl2004} and bring an additional information on random triggering probability. As these results depend on the $\delta t$ parameter, one should note that they could be improved by reducing the coincidence window, which is in principle possible due to the shortness of the analog pulses. In our case, the 3.2 GS/s sampling rate was the limiting factor since $\delta t = \SI{1.25}{\nano\second}$ only corresponds to four consecutive waveform samples. Besides, we verified that reducing $\delta t$ induced an increase of the true coincidence detection efficiency, particularly for low $V_{th}$ values. An additional event selection criterion based on time-over-threshold (TOT) could be used to reject high frequency noise-generated triggers.

\subsection{Timing performance}
\subsubsection{68 MeV protons}

Figure \ref{fig:ARRONAX2019_eff_timeRes_somme} also shows the results of the timing measurements that were carried out on the same acquired datasets. On Figure \ref{fig:ARRONAX2019_eff_timeRes_somme}, the measured TOF resolution is plotted as a function of $V_{th}$ for the three pCVD detectors (solid lines). A similar evolution of the TOF resolution can be observed with the three pCVD samples. It can be noticed that as long as the threshold level remains below the value maximizing the detection efficiency, the measured TOF resolution is rather constant. For higher values, as the threshold rejects low amplitude signals, the SNR of the selected events increases. Since the time resolution of diamond detectors is directly related to the SNR \citep{BossiniMinafra2020, Ciobanu2011}, the TOF resolution improves as well. In the case of the Element6 detector, the TOF resolution ranges from \SI{220}{\pico\second} (RMS) to \SI{162}{\pico\second} (RMS) and \SI{218}{\pico\second} (RMS) is obtained at best efficiency. The TOF resolution measured with the II-VI samples ranges from \SI{227}{\pico\second} (RMS) to \SI{172}{\pico\second} (RMS) (\SI{225}{\pico\second} at best efficiency). The DDK provides the best results with a TOF resolution ranging from \SI{192}{\pico\second} (RMS) to \SI{139}{\pico\second} (RMS) (\SI{191}{\pico\second} at best efficiency).

The overall better performance obtained with the DDK detector is related to the capacitance of the devices. That plays a crucial role in timing measurements \citep{BossiniMinafra2020, Ciobanu2011}. Using the geometries defined in Table \ref{tab:all_diamonds} and the relative permittivity of diamond ($\varepsilon_{r} = 5.7$), the DDK detector's computed capacitance is \SI{1.2}{\pico\farad} compared to $6.5$ and \SI{3.9}{\pico\farad} for the Element6 and II-VI detectors respectively. Despite that, the Element6 sample's timing response appears to be slightly better than that of the II-VI sample. It therefore tends to show the superior performance of the Element6 pCVD detector compared to the II-VI one. As a comparison, in a previous beam test, the sCVD-DOI reference pair of diamond detectors had been tested in similar conditions and reached a \SI{94}{\pico\second} ($\sigma$) TOF resolution \citep{Marcatili2020}.

\subsubsection{95 MeV/u carbon ions}

\begin{figure}[b!]
    \begin{center}
    \includegraphics[width=1.0\columnwidth]{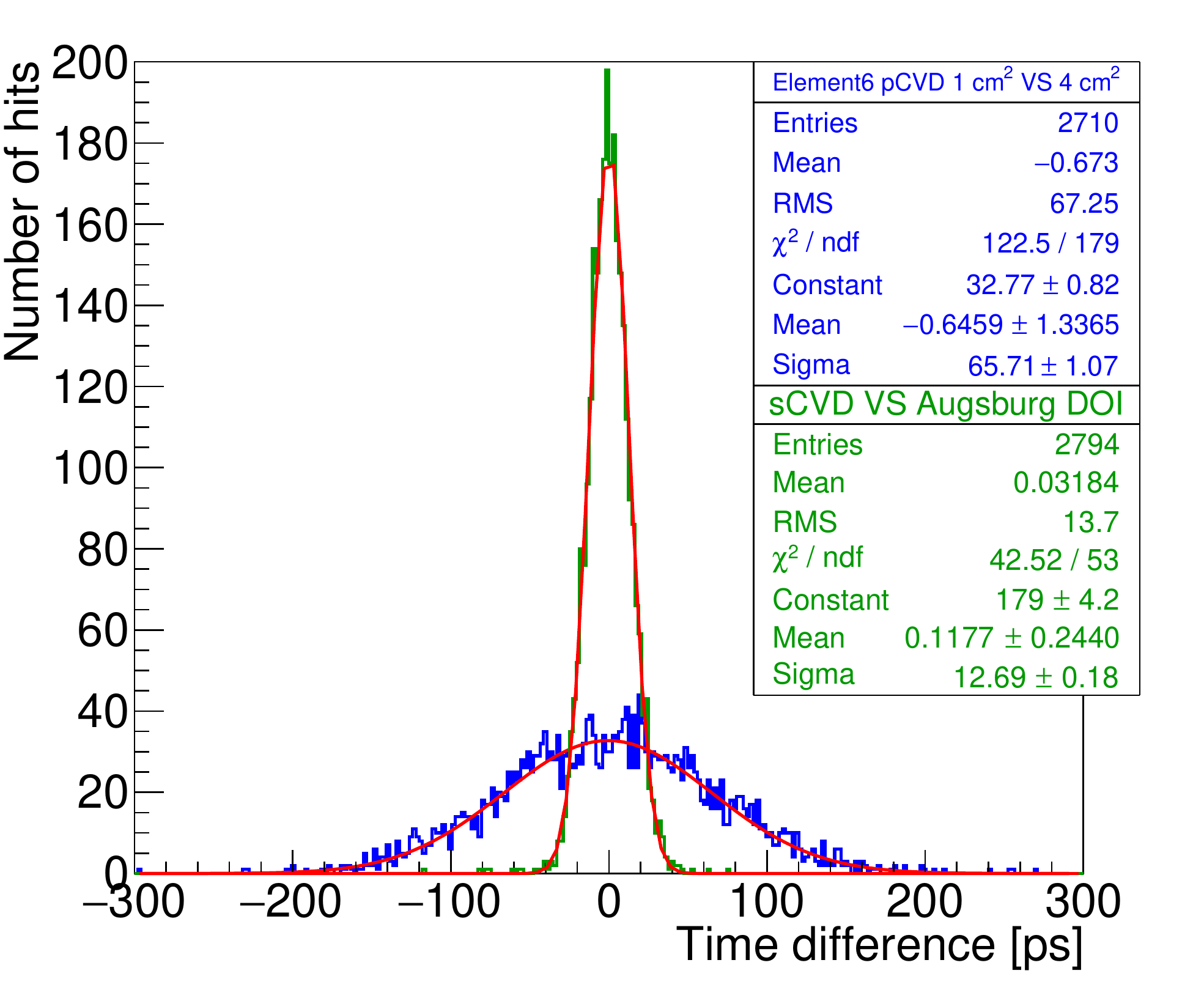}
    \caption{Time difference distributions obtained with two pairs of diamond detectors and single 95 MeV/u carbon ions at GANIL The two pairs were the sCVD-DOI reference pair (green) and two large area pCVD detectors (blue).}
    \label{fig:GANIL_timeRes}
    \end{center}
\end{figure}

The results of the timing measurements carried out at GANIL are presented in Figure \ref{fig:GANIL_timeRes}. Due to the large energy deposition in the detectors (25 MeV in DOI and 44 MeV in sCVD according to SRIM simulations \citep{Ziegler2010}), the high SNR enabled us to lower the discrimination fraction down to 20\%. Thus, the two detector pairs highlighted excellent results. In each case, the distribution could be fitted to derive the $\sigma_{TOF}$ value. The measured TOF resolution of the sCVD-DOI pair is $\sigma_{TOF} = \SI{13}{\pico\second}$. In the case of the pCVD pair, the obtained TOF resolution was \SI{66}{\pico\second} ($\sigma$). The difference between the results obtained with the two pairs can be explained by the quality of the involved samples and the large size of the pCVD detectors (with computed capacitances of \SI{20}{\pico\farad} and \SI{6.5}{\pico\farad}). In any case, the two pairs demonstrated excellent results nicely fitting with the objectives of the hodoscope.

\subsubsection{Bunches of 8.53 keV synchrotron radiation X-rays}

\begin{figure}[b!]
    \centering
  \begin{tabular}{c}
  \resizebox{0.9\columnwidth}{!}{\includegraphics{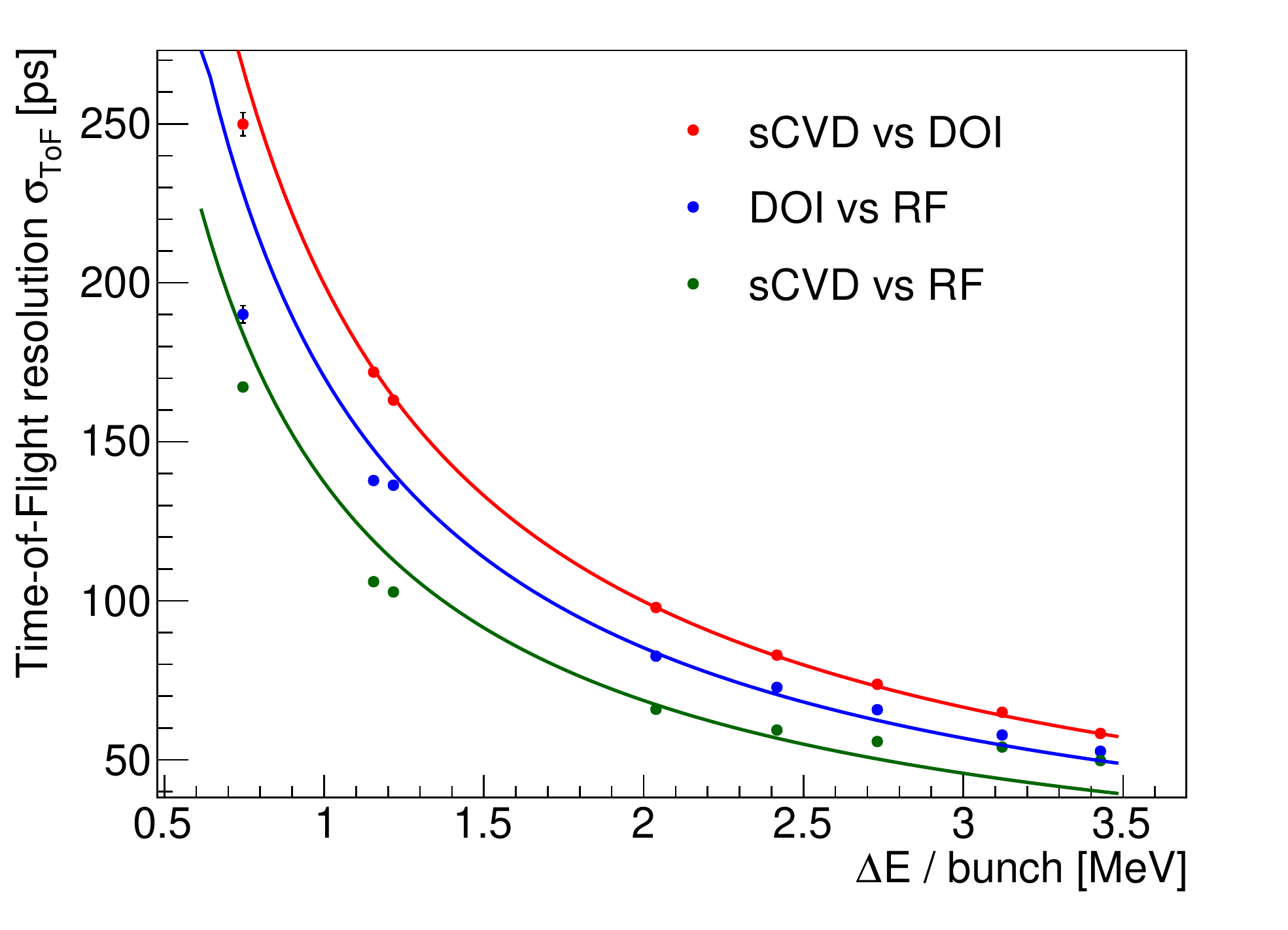}}
  \\
  \resizebox{0.9\columnwidth}{!}{\includegraphics{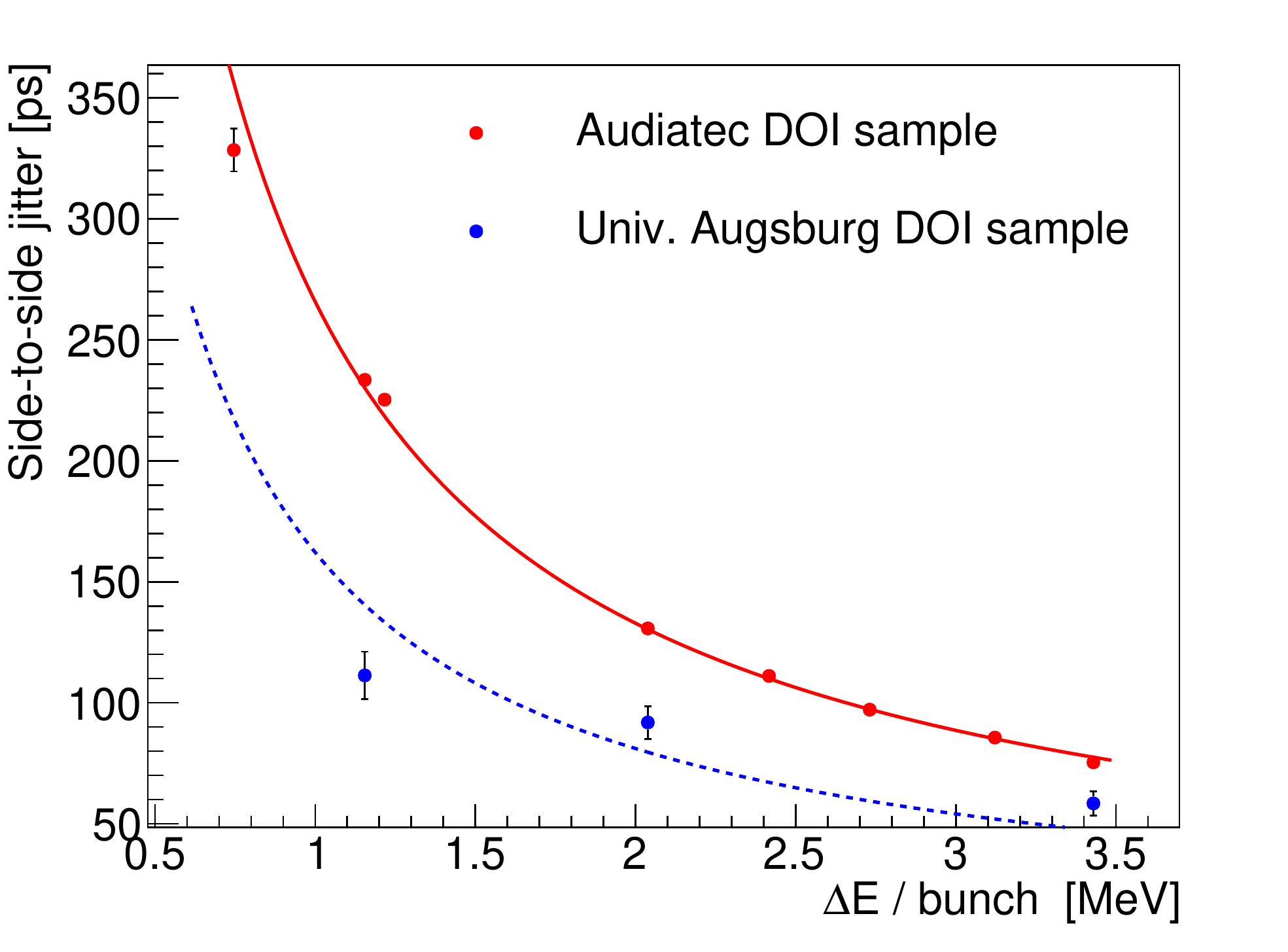}}
  \end{tabular}
    \caption{(Top) TOF resolution as a function of the energy deposited by a bunch of 8.53 keV X-rays in a pair of diamond detectors composed of a Element6 sCVD and a Audiatec DOI. (Bottom) Standard deviation of the side-to-side pulses time difference measured on the Audiatec DOI detector (red) and the Augsburg University DOI detector (blue) as a function of the energy deposition in the detectors. }
    \label{fig:ESRF_timeRes}
\end{figure}

Figure \ref{fig:ESRF_timeRes}-(Top) represents, for each energy deposition (each attenuator), the TOF resolution measured between the two detectors (red) and between each detector and the beam RF (DOI = blue and sCVD = green) as a function of the deposited energy. The results are fitted with a function $\sigma_{TOF} = C/\Delta E$ with $C$ a parameter to highlight the correlation between the TOF resolution and the deposited energy. Due to the low jitter in the beamline RF signal, the TOF measurements using the RF signal and a single diamond detector give better results than TOF measurements made between two diamond detectors. It also provides a common reference allowing us to deduce that the sCVD detector gives a better result than the Audiatec one. The results obtained with these two DOI detectors (Audiatec and Augsburg) are compared in Figure \ref{fig:ESRF_timeRes} (Bottom). As the electronic channels used in both cases were identical (1 CIVIDEC C2-HV per channel), the contribution of the electronic jitter is the same for the two detectors. The better timing response of the Augsburg DOI can therefore be related to the intrinsic better performance of the detector, in comparison with the Audiatec one. While the side-to-side jitter evolution measured on the Audiatec detector fits pretty well with an inverse function, it is not the case of the Augsburg DOI sample. The dashed line fit is drawn to show which correlation would be expected with these measurements but they seem to be less sensitive to the energy deposition. A possible explanation may rely on the surface heterogeneity of the Augsburg DOI sample, already highlighted in \citep{GallinMartel2020}, that could explain that the signal shape will depend on the hit position on the detector. Therefore, the jitter of the Augsburg sample could be dominated by other factors than the energy deposition.

\subsubsection{Minimum Ionizing Particles ($\beta$ source)}
\begin{figure}[b!]
    \begin{center}
    \includegraphics[width=1.0\columnwidth]{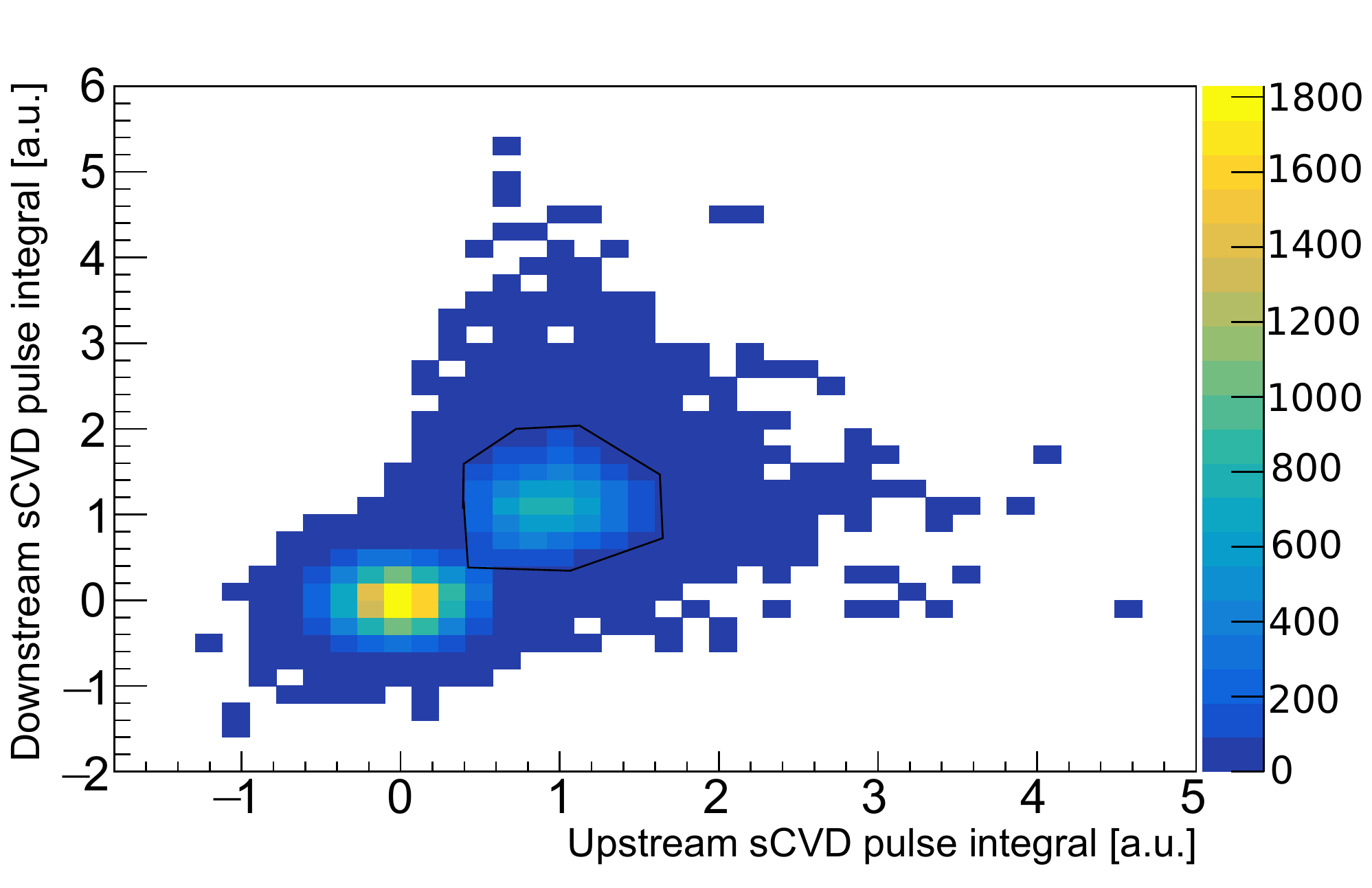}
    \caption{Correlation between the charge response of the two sCVD detectors used for the MIP timing measurement (the acquisition is triggered by the external downstream scintillator). The first peak centered on (0;0) is the noise peak. The second peak is due to single high energy electrons depositing the same amount of energy in the two detectors. The contour drawn on the distribution is the graphical cut applied on the data to measure the time resolution.}
    \label{fig:beta_graphical_cut}
    \end{center}
\end{figure}

Prior to the timing measurement itself, a preliminary analysis was performed at LPSC. As the acquisition was triggered by the downstream scintillator, it is shown in Figure \ref{fig:beta_graphical_cut} that one can assess the existing correlation between the responses of the two diamond detectors. Each detector response corresponds to the integral of the sum signals. The result mainly exhibits two distributions. The first one is centered on zero and the second corresponds to a signal measured simultaneously on both detectors. The statistical predominance of the distribution centered on zero is due to the trigger on the external scintillator which has i) a larger area than diamonds, and may then detect electrons outside the diamond active areas, and ii) a low detection threshold, enabling triggering on background. 

\begin{figure}[h!]
    \begin{center}
     \includegraphics[width=1.0\columnwidth]{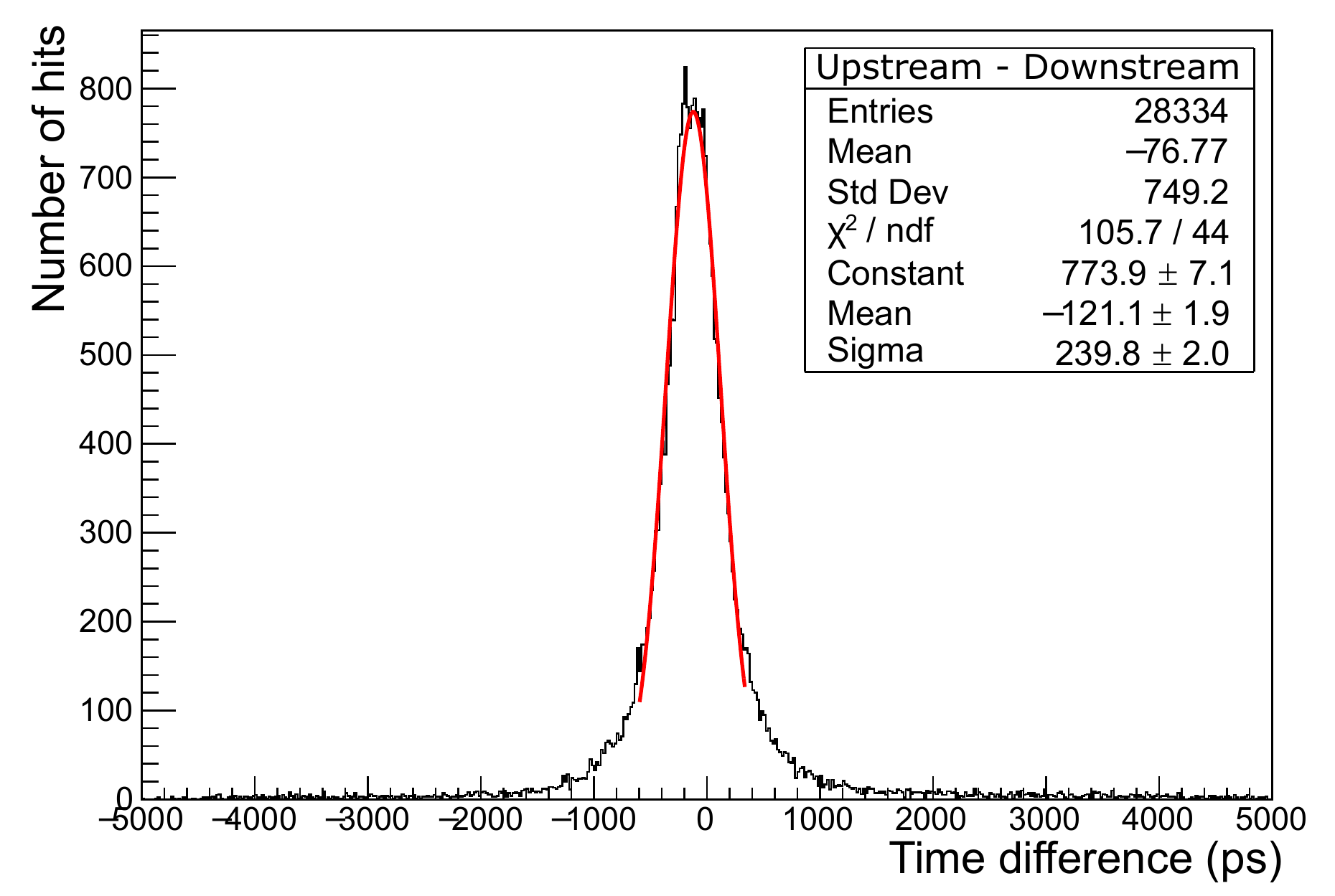}
    \caption{Distribution of the time difference between the sum signals of the two Element6 sCVD detectors detecting the same high energy $\beta$ electrons.}
    \label{fig:beta_timeRes}
    \end{center}
\end{figure}

In order to measure the time resolution of the detectors, a graphical selection was performed on the data as illustrated by the black contour drawn in Figure \ref{fig:beta_graphical_cut}. Since the electrons of highest energy are close to MIP, their energy deposition in the two detectors is expected to be almost constant. By selecting the events which exhibit the same charge response in both detectors, we can thus select electrons in the higher energy part of the beta spectrum. The time difference measured on the sum signals of the selected events is presented in Figure \ref{fig:beta_timeRes}. Different estimators can then be used to derive the TOF resolution in this case. An optimistic estimation would consist in using the standard deviation given by a Gaussian fit. Choosing such a parameter neglects the tails present on both sides of the distribution. Under these conditions, $\sigma_{TOF} = \SI{240}{\pico\second}$ is obtained, which corresponds to a timing resolution of \SI{170}{\pico\second} for a single detector. A more objective estimator is the RMS of the distribution. This one takes into account its tails that strongly degrade the TOF resolution. Using this estimator, the TOF resolution is \SI{749}{\pico\second} (RMS), \textit{i.e.} a timing resolution of \SI{530}{\pico\second} (RMS) for one detector. However, these values were obtained using a \SI{10}{\nano\second} coincidence window, which is of the same order of the signal duration, and may contain random coincidences.

\subsubsection{Summary of Time-Of-Flight measurements}

The TOF resolution we measured at laboratory, at ARRONAX and GANIL are summarized in Table \ref{tab:TOF_results} where D1 is the upstream detector and D2 is the downstream one.

\begin{table*}[t!]
\centering
\caption{Summary of the different timing measurements presented. The energy deposition of single ions has been estimated with SRIM simulations. *Result from a previous study \citep{Marcatili2020}, given here for completeness purpose.}
\label{tab:TOF_results}
\resizebox{\textwidth}{!}{%
\begin{tabular}{|c|c|c|c|c|c|c|c|c|}
\hline
\begin{tabular}[c]{@{}c@{}}Diamond\\ D1\\ VS\\ D2\end{tabular} &
  Manuf. &
  \begin{tabular}[c]{@{}c@{}}Size\\ (\si{\cubic\milli\metre})\end{tabular} &
  \begin{tabular}[c]{@{}c@{}}Computed\\ capacitance\\ (pF)\end{tabular} &
  \begin{tabular}[c]{@{}c@{}}Particle\\ type\end{tabular} &
  \begin{tabular}[c]{@{}c@{}}Particle\\ energy\\ (MeV)\end{tabular} &
  \begin{tabular}[c]{@{}c@{}}Energy deposition\\ per particle/pulse\\ (MeV)\end{tabular} &
  \begin{tabular}[c]{@{}c@{}}Sum\\ signals\\ used ?\end{tabular} &
  \begin{tabular}[c]{@{}c@{}}Measured\\ TOF\\ resolution\\ (ps $\sigma$)\end{tabular} \\ \hline
sCVD &
  \multirow{2}{*}{E6} &
  \multirow{2}{*}{4.5 $\times$ 4.5 $\times$ 0.517} &
  \multirow{2}{*}{0.7} &
  \multirow{2}{*}{\begin{tabular}[c]{@{}c@{}}$^{90}$Sr decay\\ electron\end{tabular}} &
  \multirow{2}{*}{$\sim$ MIP} &
  \multirow{2}{*}{$\sim$ 0.3} &
  \multirow{2}{*}{\ding{51}} &
  \multirow{2}{*}{240 $\pm$ 2} \\
sCVD &
   &
   &
   &
   &
   &
   &
   &
   \\ \hline
\multirow{6}{*}{\begin{tabular}[c]{@{}c@{}}DOI\\ sCVD\end{tabular}} &
  \multirow{4}{*}{\begin{tabular}[c]{@{}c@{}}Augsburg\\ E6\end{tabular}} &
  \multirow{4}{*}{\begin{tabular}[c]{@{}c@{}}5 $\times$ 5 $\times$ 0.3\\ 4.5 $\times$ 4.5 $\times$ 0.517\end{tabular}} &
  \multirow{4}{*}{\begin{tabular}[c]{@{}c@{}}1.2\\ 0.7\end{tabular}} &
  \multirow{2}{*}{proton} &
  \multirow{2}{*}{68*} &
  1.0 &
  \multirow{2}{*}{\ding{51}} &
  \multirow{2}{*}{94.1 $\pm$ 0.4*} \\
 &
   &
   &
   &
   &
   &
  1.6 &
   &
   \\ \cline{5-9} 
 &
   &
   &
   &
  \multirow{2}{*}{carbon ion} &
  \multirow{2}{*}{1140} &
  25 &
  \multirow{2}{*}{\ding{55}} &
  \multirow{2}{*}{12.7 $\pm$ 0.2} \\
 &
   &
   &
   &
   &
   &
  44 &
   &
   \\ \cline{2-9} 
 &
  \multirow{2}{*}{\begin{tabular}[c]{@{}c@{}}Audiatec\\ E6\end{tabular}} &
  \multirow{2}{*}{\begin{tabular}[c]{@{}c@{}}5 $\times$ 5 $\times$ 0.3\\ 4.5 $\times$ 4.5 $\times$ 0.517\end{tabular}} &
  \multirow{2}{*}{\begin{tabular}[c]{@{}c@{}}1.2\\ 0.7\end{tabular}} &
  \multirow{2}{*}{\begin{tabular}[c]{@{}c@{}}X-ray pulse\\ (no attenuator)\end{tabular}} &
  \multirow{2}{*}{$8.53 \cdot 10^{-3}$} &
  3.4 &
  \multirow{2}{*}{\begin{tabular}[c]{@{}c@{}}DOI\\ only\end{tabular}} &
  \multirow{2}{*}{58.3 $\pm$ 0.5} \\
 &
   &
   &
   &
   &
   &
  3.3 &
   &
   \\ \hline
\multirow{6}{*}{\begin{tabular}[c]{@{}c@{}}pCVD\\ sCVD\end{tabular}} &
  \multirow{2}{*}{\begin{tabular}[c]{@{}c@{}}E6\\ E6\end{tabular}} &
  \multirow{2}{*}{\begin{tabular}[c]{@{}c@{}}10 $\times$ 10 $\times$ 0.3\\ 4.5 $\times$ 4.5 $\times$ 0.517\end{tabular}} &
  6.5 &
  \multirow{6}{*}{proton} &
  \multirow{6}{*}{68} &
  1.0 &
  \multirow{2}{*}{\ding{51}} &
  \multirow{2}{*}{218 $\pm$ 1} \\
 &
   &
   &
  0.7 &
   &
   &
  1.6 &
   &
   \\ \cline{2-4} \cline{7-9} 
 &
  \multirow{2}{*}{\begin{tabular}[c]{@{}c@{}}II-VI\\ E6\end{tabular}} &
  \multirow{2}{*}{\begin{tabular}[c]{@{}c@{}}10 $\times$ 10 $\times$ 0.5\\ 4.5 $\times$ 4.5 $\times$ 0.517\end{tabular}} &
  3.9 &
   &
   &
  1.6 &
  \multirow{2}{*}{\ding{51}} &
  \multirow{2}{*}{225 $\pm$ 1} \\
 &
   &
   &
  0.7 &
   &
   &
  1.6 &
   &
   \\ \cline{2-4} \cline{7-9} 
 &
  \multirow{2}{*}{\begin{tabular}[c]{@{}c@{}}DDK\\ E6\end{tabular}} &
  \multirow{2}{*}{\begin{tabular}[c]{@{}c@{}}5 $\times$ 5 $\times$ 0.3\\ 4.5 $\times$ 4.5 $\times$ 0.517\end{tabular}} &
  1.2 &
   &
   &
  1.0 &
  \multirow{2}{*}{\ding{51}} &
  \multirow{2}{*}{191 $\pm$ 1} \\
 &
   &
   &
  0.7 &
   &
   &
  1.6 &
   &
   \\ \hline
\multirow{2}{*}{\begin{tabular}[c]{@{}c@{}}pCVD\\ pCVD\end{tabular}} &
  \multirow{2}{*}{\begin{tabular}[c]{@{}c@{}}E6\\ E6\end{tabular}} &
  \multirow{2}{*}{\begin{tabular}[c]{@{}c@{}}20 $\times$ 20 $\times$ 0.5\\ 10 $\times$ 10 $\times$ 0.3\end{tabular}} &
  20 &
  \multirow{2}{*}{carbon ion} &
  \multirow{2}{*}{1140} &
  44 &
  \multirow{2}{*}{\ding{55}} &
  \multirow{2}{*}{65.7 $\pm$ 1.1} \\
 &
   &
   &
  6.5 &
   &
   &
  25 &
   &
   \\ \hline
\end{tabular}%
}
\end{table*}

The correlation between time resolution and energy deposition (and therefore SNR) can be clearly observed. The measurements are better with the sCVD-DOI reference pair. Using carbon ions at GANIL, the performance of the pCVD pair is excellent. Considering the large energy deposition of carbon ions with energies in the hadrontherapy range, developing a pCVD hodoscope reaching a time resolution $\leq$ \SI{100}{\pico\second} ($\sigma$) is achievable in carbon ion therapy. Finally, the measurement with beta electrons allowed us to define a lower limit to these TOF resolutions.

\subsection{Proton counting}
\label{subsec:results_counting}

\begin{figure}[t!]
    \centering
  \begin{tabular}{c}
  \resizebox{0.9\columnwidth}{!}{\includegraphics{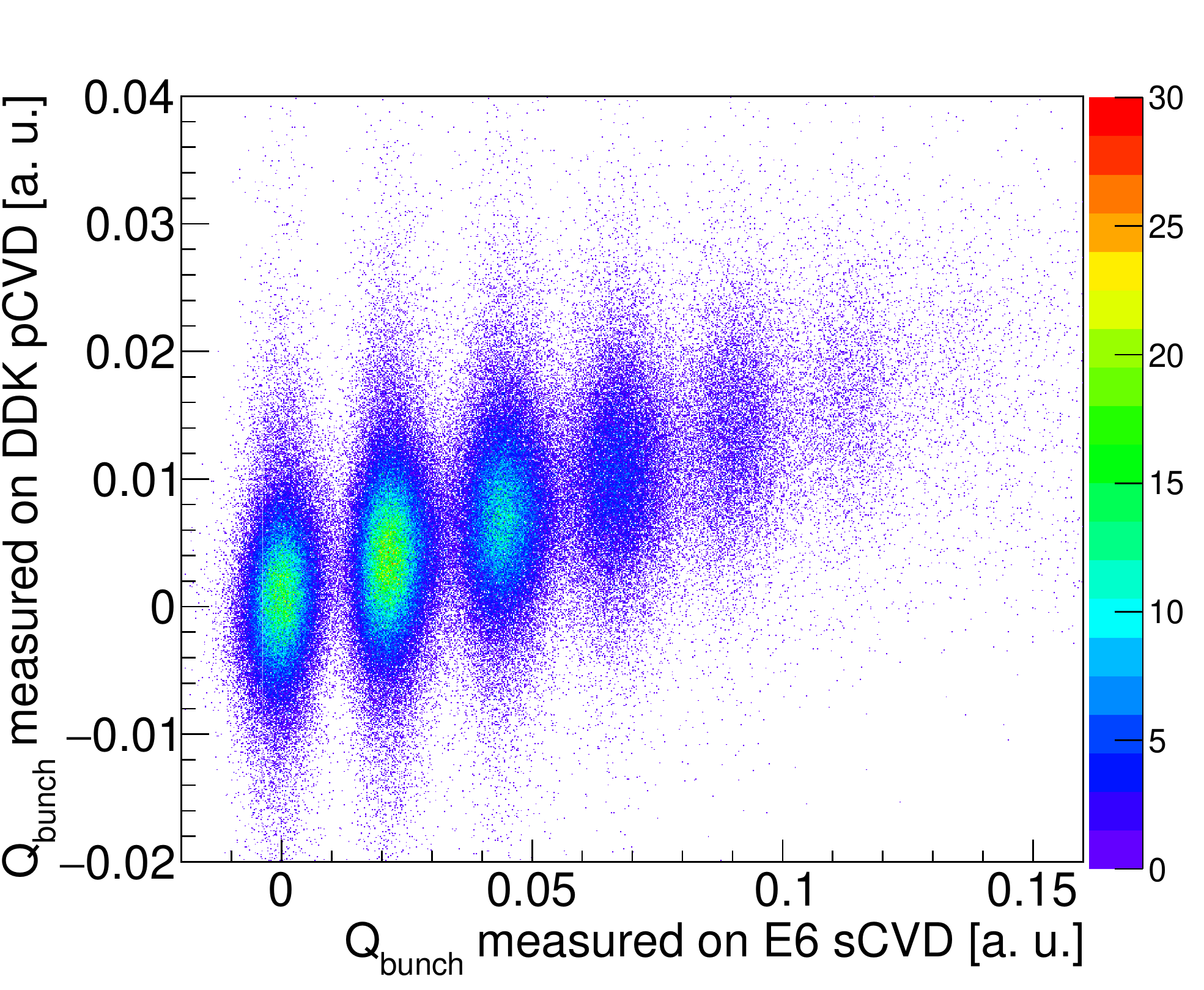}}
  \\
  \resizebox{0.9\columnwidth}{!}{\includegraphics{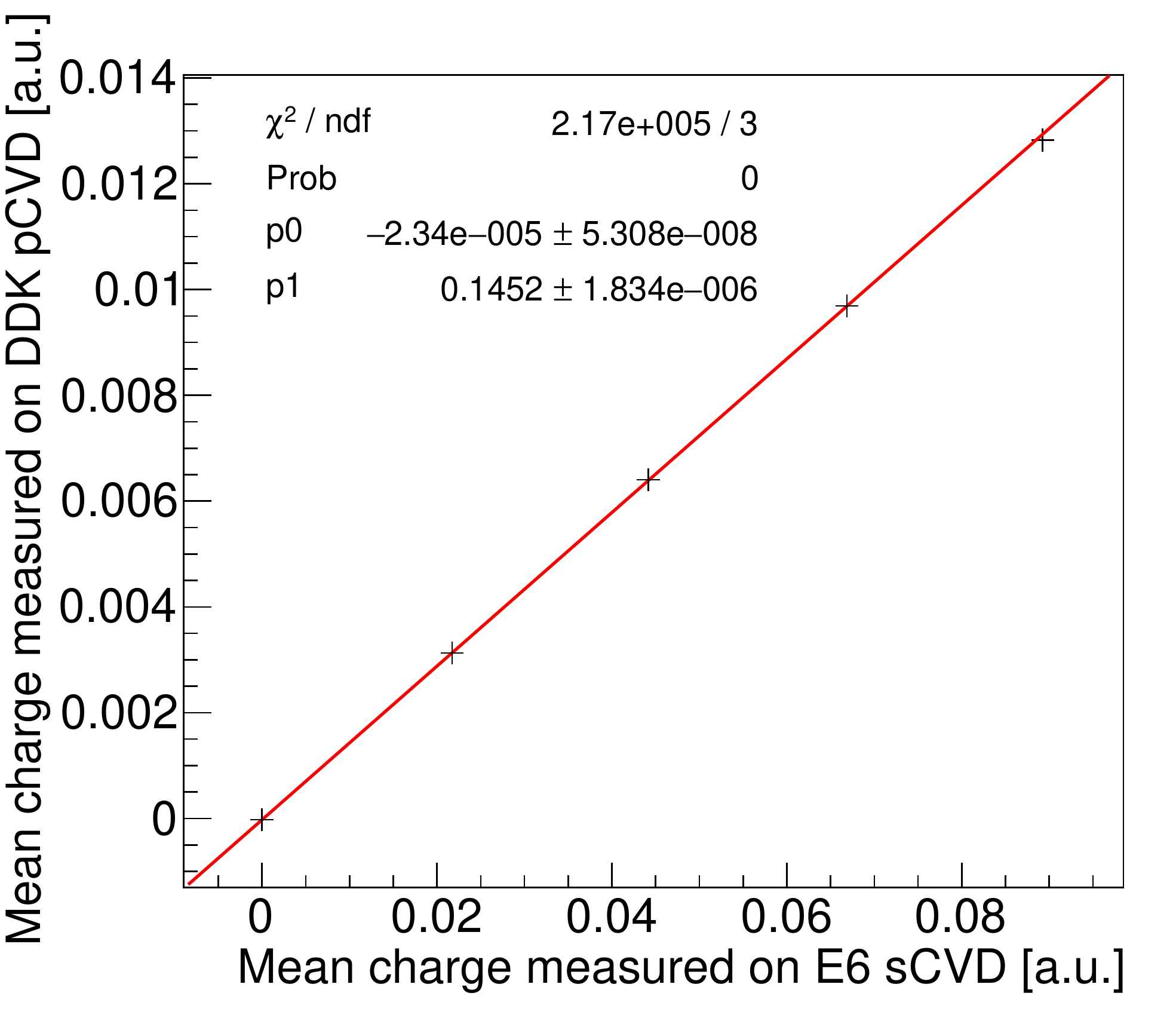}}
  \end{tabular}
    \caption{(Top) Bunch-generated ionisation charge measured on the Element6 sCVD detector as a function of the charge generated in the DDK detector. (Bottom) Mean charge generated in the sCVD detector as a function of the mean charge generated in the DDK detector, for a discrete number of protons in the bunch. Error bars are given in the figure but are hidden by the marker size. They correspond to the statistical error obtained for each number of protons.}
    \label{fig:ARRONAX_counting}
\end{figure}

The monitoring and counting capabilities of sCVD and pCVD detectors were evaluated at ARRONAX at a beam intensity around 1 proton/bunch. The results of the bunch-generated ionisation charge $Q_{bunch}$ as measured simultaneously on the sCVD and pCVD detectors are presented in Figure \ref{fig:ARRONAX_counting} (Top). On the one hand, it can be clearly observed that the sCVD detector has an energy resolution which is sufficient to distinguish a discrete number of protons contained in each bunch. The 2D distribution therefore exhibits 6 peaks corresponding to bunches whose content ranges from 0 to 5 protons. On the other hand, the pCVD detector's energy resolution is not good enough to count the number of protons in the bunch, leading to an overlap of the charge distributions corresponding to different numbers of protons. The different charge distributions could be separated in this case thanks to the correlation with the sCVD detector.

The sCVD $Q_{bunch}$ distribution which corresponds to the X-projection of the 2D histogram in Figure \ref{fig:ARRONAX_counting}-Top is the convolution of a Poisson distribution of parameter $\lambda$ with the Gaussian response function of the detector.  One can fit the whole distribution with the sum of 6 Gauss functions. From the obtained fit parameters and using the fact that $\lambda = (k+1) \cdot P(k+1)/P(k)$, one can derive the actual $\lambda$ value. From this analysis, an experimental value of $\lambda = 1.26 \pm 0.02$ is obtained, thus resulting in a mean beam current $I_{beam} = 6.16 \pm \SI{0.10}{\pico\ampere}$ (using Equations \ref{eq:current_1p_per_bunch} and \ref{eq:ibeam}). The error corresponds to the RMS of the $\lambda$ values obtained using the different $k$ values. Moreover, the $I_{beam}$ uncertainty could be easily reduced by increasing the integration time. Note that the method is only valid if the beam current is constant during the acquisition.

Nevertheless, the bunch content separation provided by the sCVD detector can be used to assess the linearity of the pCVD detector's mean charge response. Fixed thresholds can be set on the sCVD $Q_{bunch}$ distribution so that the response of the pCVD detector can be conditioned by the response of the sCVD detector. For each peak in the sCVD $Q_{bunch}$ distribution (ranging from 0 to 4 protons), the histogram of the corresponding charge measured on the pCVD is drawn and the obtained mean and RMS values are stored. The correlation between the mean responses of the two detectors for each number of protons can thus be plotted (Figure \ref{fig:ARRONAX_counting} Bottom). In spite of the poor pCVD energy resolution, one can note that its mean charge response remains linear with the number of protons contained in the bunch. There is no evidence of charge-saturation, and this suggests that pCVD detectors could be used at higher beam currents (typically clinical beam currents) to provide an efficient beam monitoring, where the proton bunch multiplicity prevents from counting the protons individually.


\section{Discussion}

At first, measurements were carried out to evaluate diamond single proton true coincidence detection efficiency, $i.e.$ the probability to detect a proton in a time coincidence window as short as possible (\SI{1.25}{\nano\second}), without triggering on the noise, in order to perform efficient TOF measurements on any incident proton. They were done using \SI{68}{\mega\electronvolt} protons in a single incident particle mode (50 fA). In this way, we could make measurements independent from the beam time structure. Three pCVD diamond sensors were tested. A proton coincidence detection efficiency $>96$\% is reached on the three diamond samples. To perform such a measurement, diamond detectors were read out on both sides which, in the case of an off-line data analysis, makes it possible to increase the SNR by a factor $\sqrt{2}$ when using identical read-out channels. If one is using this method online, particular care should be paid onto the exact synchronization and identical pulse shapes on the two readout channels. Indeed, we could observe that if a slight delay between the two signals is not corrected, the time resolution is degraded. Also, if the noise levels are different on the two signals, the noise level of the sum signal is dominated by the worse level as expected from Equation \ref{eq:N_somme_general}, which degrades the performance obtained with the best readout channel. This has an effect on both efficiency and timing resolution. In the case of a single channel reading, data analysis has shown that the signal to noise ratio is less favorable. It is obvious that in terms of efficiency, sCVD diamonds surpass the performance of pCVD but the commercially available surfaces remain small, which would imply combining several diamonds in the form of a mosaic to make a larger detector.

The purpose of the hodoscope is to detect each incident ion while ensuring intrinsic time resolution $\leq$ 100 ps. The best results were obtained with the sCVD-DOI reference pair. The TOF resolutions obtained with this pair of detectors are matching the objectives of the project, both with single protons of 68 MeV and with carbon ions of 95 MeV/u. Indeed, the proton \& PG TOF resolution obtained during a past ARRONAX experiment \citep{Marcatili2020} showed the capability of our detectors to discriminate PGs with a TOF resolution of 101 ps ($\sigma$), making techniques such as ultra-fast PGT very promising. Such results could not be obtained with pCVD detectors which exhibit a too low SNR to be able to measure an equivalent timing resolution with 68 MeV protons. Moreover, the threshold-based study of their detection efficiency and time resolution demonstrated that combining a detection efficiency $>90\%$ and a time resolution at the 100 ps level was not achievable. A noteworthy improvement of their time resolution could only be obtained for threshold values that rejected most of the single-proton signals, thus dramatically deteriorating their detection efficiency. 

It should be also considered that the energy deposition of a 68 MeV proton is the highest we can get with a single proton in particle therapy. Indeed, the protons energy range varies from 70 MeV to 250 MeV. The deposited energy, and therefore the generated signal, is inversely proportional to the proton's initial energy. The combination of these considerations makes difficult the use of pCVD detectors for time tagging of single protons in the energy range of proton therapy. We will therefore use sCVD detectors, with the limitation on the commercially available area for this application.

However, the results obtained with carbon ions at GANIL are promising. The 13 ps ($\sigma$) TOF resolution obtained between the sCVD Element6 detector and the Augsburg DOI one is the best time performance we measured, in all our experiments. This result is mainly explained by the large energy deposition generated by each ion in the diamond and by the quality of the two diamond samples. This energy deposit is so that a 66 ps ($\sigma$) resolution between two pCVD detectors was obtained whereas they were metallized with electrodes of 7 and 16 mm in diameter, respectively. Assuming that this value is the quadratic sum of their respective timing resolutions, we can estimate that their individual timing resolution is equivalent to or better than 66 ps. Besides, in the case of carbon ion therapy, the energy of the ions ranges from 95 MeV/u to 400 MeV/u. SRIM simulations show that in a 500 \si{\micro\metre} pCVD diamond with a charge collection efficiency of 30\% (measured on an alpha test bench at laboratory) generates a collected charge ranging from 156 fC to 61 fC, respectively (see \citep{Curtoni2020}). As a comparison, a 5.49 MeV $\alpha$ particle (equivalent to 67 fC) generates a sufficient signal to measure an intrinsic resolution of less than 100 ps \citep{Gallin-Martel2016, Gallin-Martel2018}. We can thus reasonably assume that to obtain an intrinsic temporal resolution of 100 ps ($\sigma$) with a large size (pCVD) detector remains a realistic goal for carbon ion therapy.

It should also be noted that in the configuration of our tests, pCVD sensors were not optimized for timing measurements. An improvement of their timing performance could be obtained by combining different approaches. First, their thickness could be reduced down to their charge collection distance so that the applied electric field can be higher. By doing so, a segmentation of the active surface would be necessary to over-compensate the increase of the capacitance. Using two layers of thin pCVD sensors would allow to improve the timing performance of the device by a factor $\sqrt{2}$ and they could be inclined with respect to the beam axis to increase their effective thickness.

\begin{figure}[t!]
    \begin{center}
    \includegraphics[width=1.0\columnwidth]{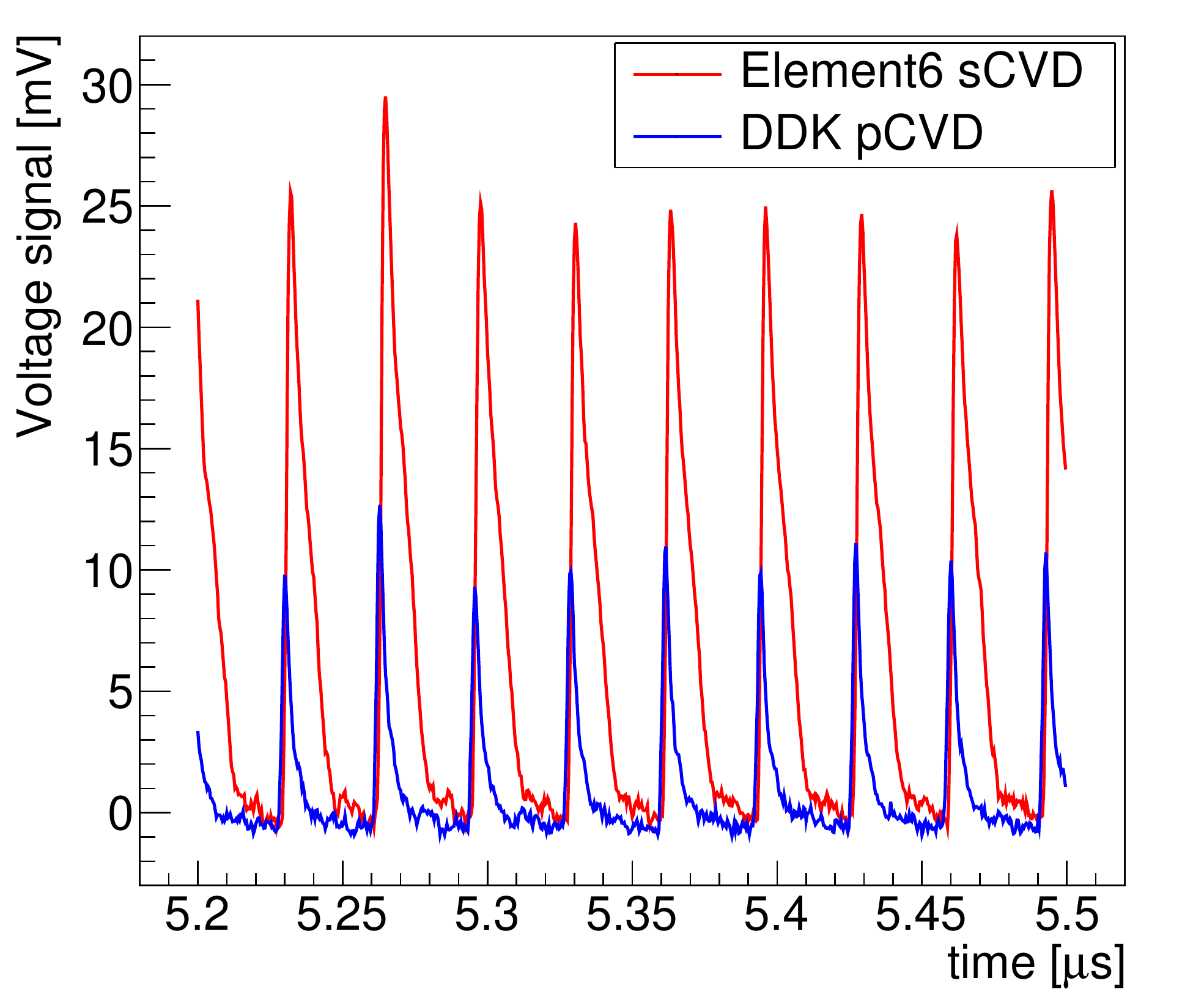}
    \caption{Compared time-domain responses of the Element6 sCVD and the DDK pCVD detectors, irradiated with the ARRONAX proton beam at $I_{beam}\sim\SI{2}{\nano\ampere}$ (with an accelerator radio-frenquency of \SI{30.45}{\mega\hertz}). The induced currents produced by the detectors are converted into voltage signals through a \SI{50}{\ohm} resistor.}
    \label{fig:WF_compared_2nA}
    \end{center}
\end{figure}

Finally, concerning the particle counting performance, the measurements carried out with 68 MeV protons at a beam intensity of $\sim$6 pA can allow us to conclude that a beam monitor equipped with sCVD diamond sensors makes it possible to provide both fast timing and counting of protons inside a bunch. In terms of hadontherapy beam monitoring, this makes it possible to count at the start of treatment at reduced beam intensity and, if necessary, identify bunches where the proton multiplicity is greater than 1. On the contrary, pCVD detectors are not able to achieve particle counting at low proton rate. This result on the comparative performance of sCVD and pCVD diamonds should however be qualified. Indeed, for higher beam intensity, sCVD diamond sensor thickness is certainly to be optimized to prevent long time drift which may result in a pile-up phenomenon at highest RF frequencies (up to 106 MHz). pCVD may present an advantage relative to sCVD. Since charge trapping occurs while charge carriers are drifting to the electrodes, it results in a shorter signal as observed in Figure \ref{fig:WF_compared_2nA} at $\sim$2 nA ($\sim$400 protons/bunch at 30.45 MHz). Such a beam current is close to clinical conditions. Therefore, the two types of diamond could be used depending on the targeted intensity range.


\section{Conclusions}

The present results are encouraging the development of a beam-tagging hodoscope with TOF capabilities. For all the tests presented in this work, sCVD diamond detectors demonstrated characteristics that are in good agreement with the requirements of the hodoscope project. The detection efficiency measurements highlighted that pCVD detectors can detect single ions with a good efficiency but can not reach a timing resolution at the order of \SI{100}{\pico\second} ($\sigma$) when detecting single protons. At low intensity, their poor energy resolution prevent them from counting the number of protons contained in a bunch but their mean charge response remains linear with the deposited energy. At higher intensity, the shorter pulses generated by pCVD detectors can represent an advantage over sCVD for beam monitoring at $100$ MHz rates. Using carbon ions, both sCVD and pCVD demonstrated excellent timing results.

Consequently, two solutions can be foreseen for the beam tagging hodoscope design. The first one may consist in using either four 4.5 $\times$ 4.5 $\times$ 0.5 \si{\cubic\milli\metre} commercially available sCVD diamonds arranged in mosaic or, later on, large area sCVD diamonds. The second solution may consist in using 20 $\times$ 20 $\times$ 0.3 \si{\cubic\milli\metre} pCVD mainly dedicated for carbon ion therapy applications. In both cases, the hodoscope will be made out of double-sided strip sensors. It will provide the ion transverse position with a precision $\leq \SI{1}{\square\milli\metre}$ (X and Y strips width). The influence of the segmentation of the metallic contacts on the timing performance of the device will have to be evaluated while it will not be possible to use the side-to-side signal summation method. The next step of the hodoscope development is the assembling of the two selected diamonds types with front-end electronics currently developed at LPSC for TOF measurements of prompt-gamma in view of range verification in particle therapy.

\section*{Acknowledgements}

The authors would like to acknowledge the ESRF-ID21 beamline for provision of synchrotron radiation with experiments MI-1243 (2016) and MI-1285 (2017), and support from the ESRF BCU group for integrating the triggered readout of the LeCroy DSO into the ID21 SPEC data acquisition system. This work was supported by Plan Cancer (CLaRyS-UFT project), the LabEx PRIMES (ANR-11-LABX-0063), FranceHadron (ANR-11-INBS-0007) and ANR MONODIAM-HE (ANR-089520). The cyclotron Arronax is supported by CNRS, Inserm, INCa, the Nantes University, the Regional Council of Pays de la Loire, local authorities, the French government and the European Union. This work has been, in part, supported by a grant from the French National Agency for Research called “Investissements d'Avenir”, Equipex Arronax-Plus noANR-11-EQPX-0004, Labex IRON noANR-11-LABX-18- 01and ISITE NExT no ANR-16-IDEX-007. It was performed in the frame of ENSAR2/MediNet network (Horizon2020-654002). The authors are grateful to Matthias Schreck from Augsburg University and Martin Fischer from Audiatec Augsburg for providing the LPSC laboratory with DOI samples. Dominique Breton and Jihanne Maalmi from IJC-Lab Orsay and Eric Delagnes from CEA Saclay are thanked for their implication in dedicated software development and technical support of the WaveCatcher data acquisition system. SC, MLGM, AB, JC, DD, LGM, AG, AL, SM, OR, FER, and MY are members of the RD42 collaboration at CERN. The authors would like to thank the reviewer for his/her useful discussion about the enhancement of the timing performance of pCVD sensors that improved the quality of the discussion.

\label{}




\bibliographystyle{elsarticle-num} 
\bibliography{library}





\end{document}